\documentstyle[]{fbssuppl}
\input psfig.sty

\title{Double Pion Production Reactions}
\author{E. Oset\instnr{1}, J. A. G\'omez Tejedor\instnr{2}, F. Cano\instnr{3}, 
J.C. Nacher\instnr{1}, S. Kamalov\instnr{4}, L. Alvarez-Ruso\instnr{1} and
E. Hern\'andez\instnr{5}}
\instlist{Departamento de F\'{\i}sica Te\'orica and IFIC, Centro Mixto 
Universidad de Valencia - CSIC, 46100 Burjassot (Valencia), Spain \and
Departamento de F\'{\i}sica Aplicada, Universidad Polit\'ecnica
de Valencia, Valencia, Spain \and
Dipartimento di F\'{\i}sica, University of Trento, Trento, Italy \and
Institut f\"{u}r Theoretische Physik, Universit\"{a}t Mainz, Germany \and
Departamento de F\'{\i}sica, Universidad de Salamanca, Spain}

\begin{document}
\maketitle

\begin{abstract}
We report on reactions producing two pions induced by
real  and virtual photons or nucleons. The role of different 
resonances in these reactions is emphasized. Novel results on
coherent $2 \pi$ photoproduction in nuclei are also reported.
\end{abstract}

\section{Double pion photoproduction on the nucleon.}

The  $\gamma N \rightarrow \pi^+ \pi^- N$ reaction is attracting attentions
of both theoretical and experimental groups and is bound to play a
significant role in photonuclear reactions much as the
$\gamma N \rightarrow \pi N$ played in the past.

Apart from the work of the Valencia group which I will report here, there
is work by other groups. A simplified model containing many of the 
important features of the reaction was worked out in \cite{a1} and
improved in \cite{a2}, \cite{1}. The model of \cite{1} contains more
mechanisms than the one of \cite{a2}, presently under revision
\cite{a3}. On the other hand, the model  of \cite{a2} incorporates an 
approximate unitarization prescription which allows one to go to higher 
energies with the model. In ref. \cite{a4} a simplified model is also used
incorporating, however, some $\rho$ decay channels. This work  has been
revised in \cite{a5} in view that some mass distribution was in
disagreement with the data, and a new parametrization is offered, which
relies on a range parameter for the $\rho$ of the order of 
$200 \, MeV$, which would not accommodate easily other known
facts of phenomenology as the isovector $\pi N$ s-wave amplitude.

The model of \cite{1} contains parameters determined solely from $\gamma$
and $\pi$ couplings to nucleons and resonances plus known properties
of resonance decay with some undetermined  sign borrowed from
quark models.

The $(\gamma, \pi \pi)$ reaction has also been studied at threshold with
the aim of testing chiral perturbation theory \cite{a6,a7}, 
particularly the $\gamma p \rightarrow \pi^0 \pi^0 p$  reaction where
chiral loops are very important.

The $\gamma p \rightarrow \pi^+ \pi^- p$ reaction  was studied in ref.
\cite{1} using effective Lagrangians, which incorporate the couplings
of the photon and pion to the nucleon and resonances. The $N$
and the $\Delta (1232), \, N^* (1440)$ and  $N^* (1520)$ (or $N'^*$) 
resonances were taken into account.
Furthermore, the $\rho$ as an intermediate state coupling to two pions
was also considered. The model reproduces fairly well the experimental
cross section \cite{2}. The model is further improved \cite{3} to account
for $s - d$ waves in the $N'^* \rightarrow \Delta \pi$ decay, while
at the same time reduces from 67 to 20 the number of Feynman diagrams
 needed to study the reaction in the range of Mainz energies
$E_\gamma \leq 800\; MeV$. In ref. \cite{3} this simplified model is
used to evaluate cross sections  for all other charge channels:
$\gamma p \rightarrow \pi^+ \pi^0 n \, , \,\;  \gamma p \rightarrow
 \pi^0  \pi^0 p \, , \,\; \gamma n \rightarrow \pi^+ \pi^- n \, , \; \,
\gamma n \rightarrow \pi^- \pi^0 p \, , \,\; \gamma n \rightarrow 
\pi^0 \pi^0 n  $.
The agreement with the data is overall good but some discrepancies
remain in the peak of the $\gamma p \rightarrow \pi^+ \pi^0 n$ 
reaction and its charge conjugate one, the $\gamma n \rightarrow
\pi^- \pi^0 p$ reaction, recently measured \cite{4}.

The relevance of this reaction for the resonance field is the novel
information that it provides  on the $N^* (1520)$ resonance, which
I try to explain here. In Fig. 1a, I show the  dominant diagram in the 
$\gamma p \rightarrow\pi^+ \pi^- p$ reaction. It  is the 
$\Delta N \pi \gamma$ Kroll Ruderman or gauge term. On the other hand in 
Fig. 1b, I show a diagram where the 
$N^* (1520,J^\pi = \frac{3}{2}^-)$ is photoexcited from the nucleon and
then it decays into $\Delta \pi$, the $\Delta $ decaying later into $N \pi$.

\begin{figure}[h]
\centerline{\protect\hbox{
\psfig{file=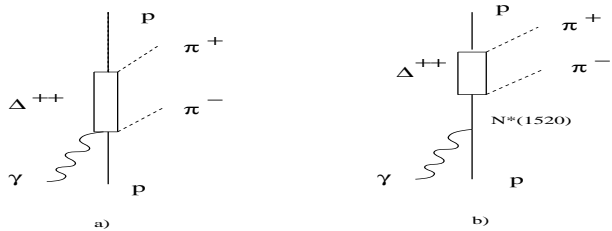,height=3.0cm,width=8.0cm}}}
\caption{Feynman diagrams}
\label{fig1}
\end{figure}

\noindent
From the 1/2 and 3/2 experimental
$N^* (1520)$ helicity amplitudes we can construct an
effective Lagrangian from where we obtain a transition operator
given by

\begin{equation}
- i \delta H = i g_\gamma \vec{S}\,^\dagger \vec{\epsilon}
 + g_\sigma
(\vec{S}\,^\dagger \times \vec{\sigma}\,) \vec{\epsilon} \ ,
\end{equation}

\noindent
where $S^\dagger $ is a spin transition operator from spin 1/2 to spin
3/2. Furthermore, we write the $N'^* \rightarrow \Delta \pi$ 
transition operator as

\begin{equation}
- i \delta H = - [\tilde{f} + \frac{\tilde{g}}{\mu^2} (\vec{S}\,^\dagger
\vec{q}\;) \; (\vec{S} \vec{q} \,)\; ] T^{\dagger \lambda} \quad + \;h. c. \ , 
\end{equation}

\noindent
where $T^\dagger $ is the isospin 1/2 to 3/2 transition operator and
$\mu$ the pion mass.
The choice of eq. (2) is not arbitrary. It allows $N'^*  \rightarrow
\Delta \pi$ decay in $s$ and $d$ waves and provides a $q$ dependence
of the amplitudes ($q$ is the CM pion momentum) which
provides the best agreement with experiment. By means of eq. (2) we can 
write the $s$ and $d$  wave decay amplitudes in 
$N'^*  \rightarrow \Delta \pi$. We find

\begin{equation}
\begin{array}{l}
A_s = - \sqrt{4 \pi} (\tilde{f} + \frac{1}{3} \tilde{g} \frac{\vec{q}\,^2}{
\mu^2}) \ ,\\[2ex]
A_d = \frac{\sqrt{4 \pi}}{3} \; \tilde{g}\; \frac{\vec{q}\,^2}{\mu^2} \ ,
\end{array}
\end{equation}

\noindent
and the width is given by

\begin{equation}
\Gamma_{N'^*  \rightarrow \Delta \pi} = \frac{1}{4 \pi^2} \frac{m_\Delta}{
m_{N'^*}} \, q (|A_s|^2 + | A_d|^2) .
\end{equation}

From the analysis of the $\pi N \rightarrow \pi \pi N$ reaction
of ref. \cite{5} one has information on $\Gamma_s , \Gamma_d$ plus
also another   ingredient, the relative
sign of $A_s$ to $A_d$ which is positive. With this information we obtain 
$A_s$
and $A_d$ up to a global sign (a sign relative to the $\gamma N \rightarrow
N^*$ amplitudes).
This sign is the first novel thing that the $\gamma p \rightarrow
\pi^+ \pi^- p $ reaction provides. Indeed we can see in Fig. 2 the
results with the two signs 
and we observe that while one of the signs is in good agreement
with the experiment (a), the other choice leads to unacceptable results (b).

\begin{figure}[h]
\centerline{\protect\hbox{
\psfig{file=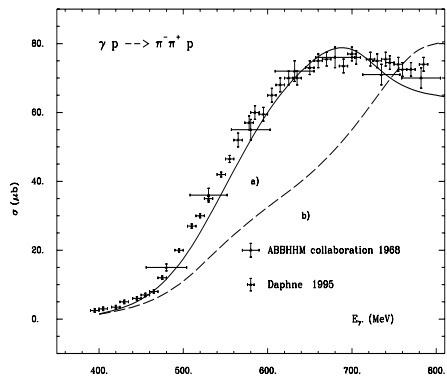,width=6.0cm,height=5.0cm,angle=-90}}}
\caption{Total cross section for $\gamma p\rightarrow \pi^+ \pi^- p $ reaction 
for different global sign}
\label{fig2}
\end{figure}

The reason for the so different results with the two signs is that the
$N^* (1520)$ mechanism of Fig. 1b interferes with the dominant one of Fig.
1a. The two amplitudes (by taking the s-wave part of the $N'^* 
\rightarrow \Delta \pi$ decay) have the same momentum and spin structure,
and the $N^* (1520)$ piece can be accounted for by making
a simple substitution in the $N \Delta \pi \gamma$ Kroll Ruderman piece:

\begin{equation}
e \frac{f^*}{\mu} \rightarrow - (g_\gamma - g_\sigma)
(\tilde{f} + \frac{1}{3} \tilde{g} \frac{\vec{q}\,^2}{\mu^2}\;)
D_{N'^*} (s) \ ,
\end{equation}

\noindent
where $D_{N'^*}$ is the $N^* (1520)$ propagator. We can see that with the
value $g_\gamma - g_\sigma = 0.157 >0$ and $\tilde{f} + \frac{1}{3} 
\tilde{g} \frac{\vec{q}\,^2}{\mu^2} > 0$ one gets a constructive
interference before the $N'^*$ pole and  a destructive one after it.
This is what can be observed in Fig. 2.

I shall not discuss here the other channels. Some results and comments
can be found in the talk of Krusche in this Workshop \cite{6}. The
$\gamma p \rightarrow \pi^0 \pi^0 p$ is well reproduced and here the 
$N^* (1520)$ term does not show up through the interference but
as the main term by itself. On the other hand there are some discrepancies
in the $\gamma p \rightarrow \pi^+ \pi^0 n$ channel which  we can
not explain so far.

\section{Repercussion of the $N^* (1520)$ findings on quark models.}

With the values of $\tilde{f} = 0.911$ and $\tilde{g} = - 0.552$ 
obtained from a fit to the $s$ and $d$ wave partial decay widths of 
$N^* (1520)  \rightarrow \Delta \pi$ and the global sign given by the
$\gamma p \rightarrow \pi^+ \pi^- p$ experiment, the amplitudes 
$A_s, A_d$ of eqs. (3) provide a definite $q$ dependence of these amplitudes.

As mentioned, this $q$ dependence is the one providing an optimal fit
to the experiment. We have checked that any other $q$ dependence of the
s-wave amplitude, consistent with the value for the on shell decay
width, provides a worse agreement.

At this point it is worth mentioning the repercussion of these results
in the quark models. This  has been shown recently \cite{8}. In this work
a nonrelativistic constituent quark model using the input of Badhuri's
model \cite{9}, adapted by Silvestre-Brac to the baryonic sector
\cite{10}, is employed, and the decay amplitudes $B \rightarrow B' \pi$
are evaluated. For this purpose one starts with a coupling of  pions
to  quarks

\begin{equation}
H_{q q \pi} = \frac{f_{qq \pi}}{\mu} \bar{\psi}_q \gamma^\mu \gamma_5
\vec{\tau} \psi_q \partial_\mu \vec{\phi}
\end{equation}

\noindent
and makes a nonrelativistic expansion keeping recoil terms 

\begin{equation}
H_{q q \pi} = f_{q q \pi} [ \vec{\sigma} \vec{q} e^{- i \vec{q} \vec{r}}
- \frac{\omega_\pi}{2 m_q} \vec{\sigma} (\vec{p} e^{-i \vec{q} \vec{r}}
+ e^{- i \vec{q} \vec{r}} \vec{p} ) ] .
\end{equation}

Now, when evaluating the $N^* , \Delta$ transition matrix elements,
since one has a radial excitation in the $N^*$ state, one needs to expand the
exponential in the first term of eq. (7) (direct term) up to order
$\vec{q}$. On the other hand, the second term of eq. (7) (recoil term) 
already gives a contribution keeping the unity in the expansion of the
exponential. Hence, we find 

\begin{equation}
\hbox{DIR} \; \propto \; \vec{q}\,^2 \quad ; \quad \hbox{REC} \; \propto
 \; 1 
\end{equation}
A direct evaluation of the $s$ and $d$-wave amplitudes
for the $N'^* \rightarrow \Delta \pi$ decay gives

\begin{equation}
\frac{A_d}{A_s} = \frac{\hbox{DIR}}{ 2 \hbox{REC} - \hbox{DIR}} \ ,
\end{equation}

\noindent
which implies

\begin{equation}
\begin{array}{rl}
A_d \, \propto & \hbox{DIR} \, \propto \; \vec{q}\,^2\\[2ex]
A_s + A_d \, \propto & \hbox{REC} \; \propto \; 1
\end{array}
\end{equation}
Hence, the non relativistic constituent  quark model keeping recoil terms
makes very clear predictions on the $q$ dependence 
of the amplitudes. Now, by looking at the $q$ dependence
demanded by the $\gamma
p \rightarrow \pi^+ \pi^- p $ reaction, expressed in eq. 3, we obtain

\begin{equation}
A_d = \frac{\sqrt{4 \pi}}{3} \tilde{g}\,  \frac{\vec{q}\,^2}{\mu^2} \;
; \, A_s + A_d = - \sqrt{4 \pi} \, \tilde{f} \ ,
\end{equation}

\noindent
which is the exactly the $\vec{q}$ dependence provided by the quark
model with recoil terms, eq. (10).

The global sign of these amplitudes prefered by the $\gamma p \rightarrow
\pi^+ \pi^- p$ experimental data is also the one provided by the
quark model. This is 
another accomplishment of these quark models, but one should recall
that not all variants of nonrelativistic, or relativized quark models
will satisfy these new constraints. This is important to note since
problems still remain when one comes to absolute values of these
amplitudes\cite{8}. Extra work is needed to explain these discrepancies, but 
it is important that these  improvements are done respecting the new 
constraints found thanks to the $\gamma p \rightarrow \pi^+ \pi^- p$ reaction. 
Actually, a treatment similar to the present one but making an expansion in 
terms of  $\omega_\pi/ E_q$ instead of $\omega_\pi / m_q$ seems
to lead to very much improved results, while keeping the consistency
with the findings discussed in this section \cite{11}.

\section{Meson exchange current and coherent $2 \pi$ photoproduction}

Assume the $\gamma N \rightarrow \pi \pi N$ reaction occurs inside
a nucleus and one of the pions, say the $\pi^-$, is produced off
shell and absorbed by a second nucleon. One obtains then meson
exchange current mechanisms which contribute to the ($\gamma, \pi^+$)
reaction in nuclei and which would be represented by diagrams like those in
Fig. 1 with the $\pi$ line attached to a nucleon line. This 
mechanism has already been explored in \cite{b1} where it was found
to contribute significantly to the $\gamma \; ^3He \rightarrow t \; \pi^+$
reaction at large momentum transfer.

In addition, the coherent $2 \pi$ photoproduction process in nuclei
has been studied in \cite{b2} and has shown very interesting features
tied to the isospin structure of the amplitudes. A photon coupling to
a nucleon can have an isoscalar and isovector component. Assume we
have the coherent reaction occurring in isospin $I = 0$ nuclei

\begin{equation}
\begin{array}{ll}
\gamma + A \rightarrow  & \pi^+ \pi^- \quad + A_{g.s.}\\
& \pi^0 \pi^0 \quad \; + A_{g.s.}
\end{array}
\end{equation}

\noindent
and let us take the isoscalar part of the amplitude. This will force
the $\pi^+ \pi^-  (\pi^0 \pi^0)$ system to have $I = 0$ and, because
of symmetry, even angular momentum $L = 0, 2 \ldots$ The isovector part
will force the $\pi^+ \pi^- (\pi^0 \pi^0)$ system into $I = 1$. This
is forbidden for the  $\pi^0 \pi^0$ system, so only the
$\pi^+ \pi^-$ can go with $I = 1$, which forces $L = 1, 3 \ldots $ The
dynamics of the elementary reaction is such that the $\gamma
N \rightarrow \pi^+ \pi^- N$ reaction is dominated by the diagram of Fig. 1a, 
where the photon behaves as an isovector, while this mechanism is forbidden for
$\pi^0 \pi^0$ production. Obvious consequences of that  are that
the $\pi^+ \pi^-$ system is largely suppressed when the pions travel
together ($L = 0$ and hence $I = 0)$.

Similarly the $\pi^0 \pi^0$ system is only produced in $I = 0$ and
hence the pions prefer to travel together. On the other hand the 
strength of the isoscalar part of the $\gamma N \rightarrow \pi \pi N$
amplitude is much smaller than the isovector part in the model of \cite{3}
and the consequence of it is that the maximum of the $\pi^0 \pi^0$
cross section is about three orders of magnitude smaller than the
maximum of the $\pi^+ \pi^-$ one. These are very strong tests of the
model which should encourage the experimentalists to perform such 
reactions.

\section{Two pion electroproduction.}

The model of ref. \cite{2} can be extended to virtual photons coming from the
$(e, e')$ vertex. These reactions are presently under experimental
investigation at TJNAF \cite{d1,d2}. We have studied the $2 \pi$
production processes where there is a $\Delta \pi$ in the final
state , ie. $e p \rightarrow e'  \Delta^{++} \pi^-$ and 
$e p \rightarrow e'  \Delta^0 \pi^+$ (with $\Delta^0 \rightarrow \pi^- p)$.
Only 8 diagrams of the model of \cite{3} contain a $\Delta$ in the final 
state and we depict these diagrams below in Fig. 3

\begin{figure}[h]
\centerline{\protect\hbox{
\psfig{file=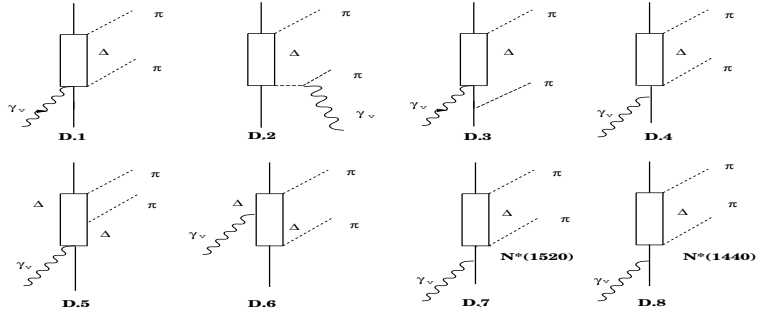,width=10.0cm,height=4.5cm}}}
\caption{Feynman diagrams used in the model for $\gamma_v p\rightarrow 
\pi\Delta$}
\label{fig3}
\end{figure}

The evaluation of the amplitudes for these reactions requires the 
extension of the model of \cite{3} to account for the zeroth component
of the electromagnetic current and the implementation of adequate form
factors. This task has been undertaken in \cite{d3}. In Fig. 4 we show
the results obtained for the cross section of the virtual photons,
defined in the standard way 

\begin{equation}
\frac{d \sigma}{d \Omega'_e d E'_e} = \Gamma (\sigma_{\gamma_v}^T +
\epsilon \sigma^L_{\gamma_v}) = \Gamma \sigma_{\gamma_v} \ ,
\end{equation}

\noindent
with $\Gamma$ and $\epsilon$ the flux factor and the polarization
of the virtual photon \cite{d4}. The results are shown
for the $e p \rightarrow e'  \Delta^{++} \pi^-$ reaction as a 
function of $Q^2 = - q^2$ and averaged over the range 
$0.3 < Q^2 < 1.4 \, GeV^2$ in order to compare with the data.
As one can see, the agreement is fair but more precise data are 
expected to come soon which will impose stronger constraints on the
theory.

\begin{figure}[h]
\centerline{\protect\hbox{
\psfig{file=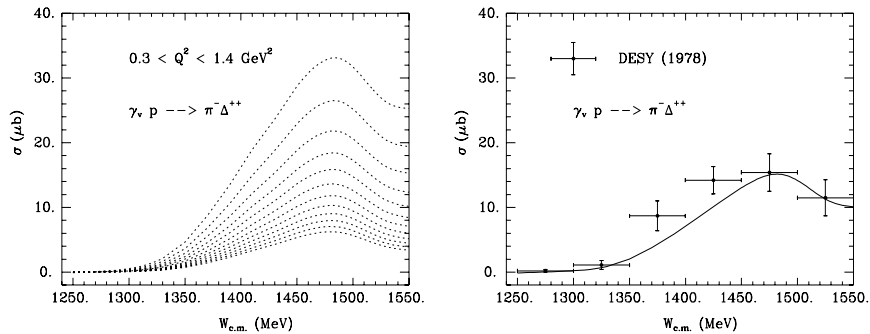,height=5cm,width=11.5cm,angle=-90}}}
\caption{Cross sections for $\gamma_v p\rightarrow\Delta^{++} \pi^-$ as a
function of the $\gamma_v p$ center of mass energy}
\label{fig4}
\end{figure}

\section{Application of isoscalar $N^*$ excitation in the $NN \rightarrow
NN \pi \pi $ reaction.}

We have developed recently a model  for the $NN \rightarrow NN \pi \pi$
reaction which contains terms coming from chiral Lagrangians, $\Delta$
excitation and Roper excitation \cite{e1}. The model is depicted
in Fig. 5. The excitation on the second nucleon and antisymmetry are 
incorporated in addition. Summarizing the results we find that the
$N^*$ excitation terms  (4 - 7), where the $N^*$ decays into 
$N (\pi \pi)^{I= 0}_{s-wave}$ are largely dominant close to threshold 
in the channels where
the two pions can be in $I = 0$ in the final state, like the 
$p p \rightarrow p p \pi^+ \pi^-$ reaction. On the other hand in the
$p p  \rightarrow p n \pi^+ \pi^0$ reaction  the $\Delta \Delta$ excitation
terms are the most important. The comparison of these two channels allows
us to appreciate the role played by $N^*$ excitation in some of the isospin
channels which, as we can see, is essential to understand the
experiment at low energies. In Fig. 6, we show the cross sections for
the $  p p  \rightarrow p p \pi^+ \pi^-$ and $p p \rightarrow p n \pi^+ 
\pi^0$ reactions. The calculations are done with plane
waves, but the results are increased at lower energies when final state
interaction is considered, and the agreement with experiment is improved.
In the figures, the total cross sections are given by the solid lines,
corresponding to two different options for the $N (\pi \pi)^{I= 0}_{s-wave}$ 
decay\cite{e1}.

\begin{figure}[h]
\centerline{\psfig{file=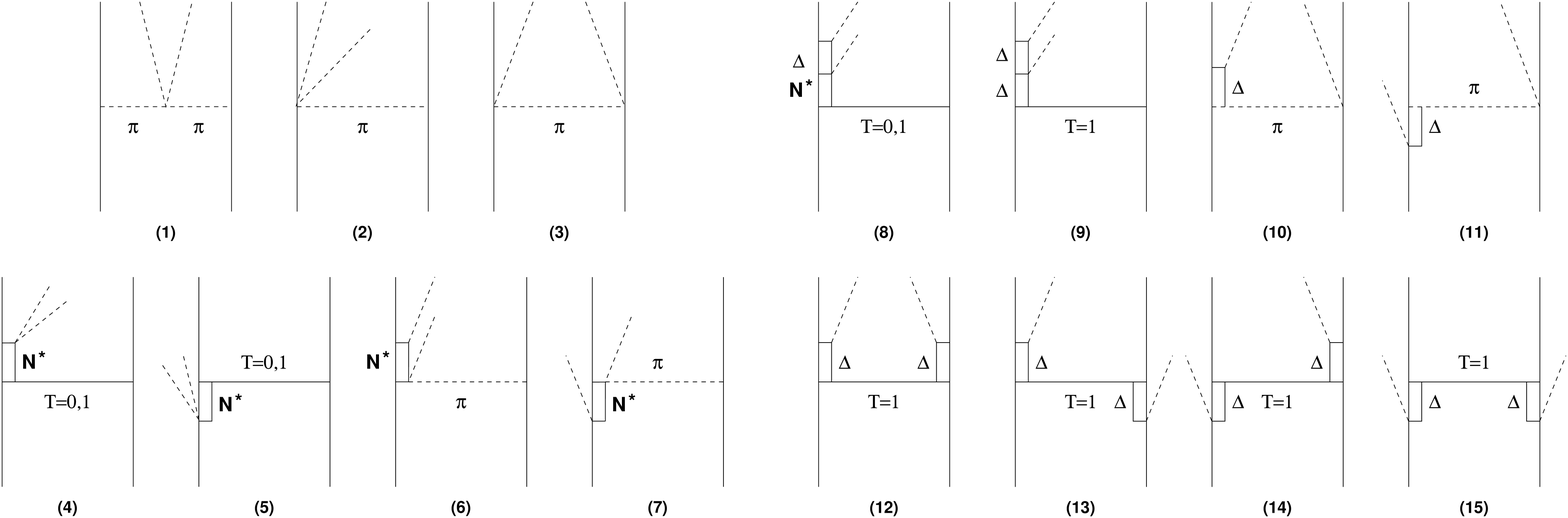,height=4.5cm,width=12.cm}}
\caption{Complete set of Feynman diagrams of our model.}
\label{fig5}
\end{figure}

\begin{figure}[h]
\begin{minipage}{.47\linewidth}
\centerline{\psfig{file=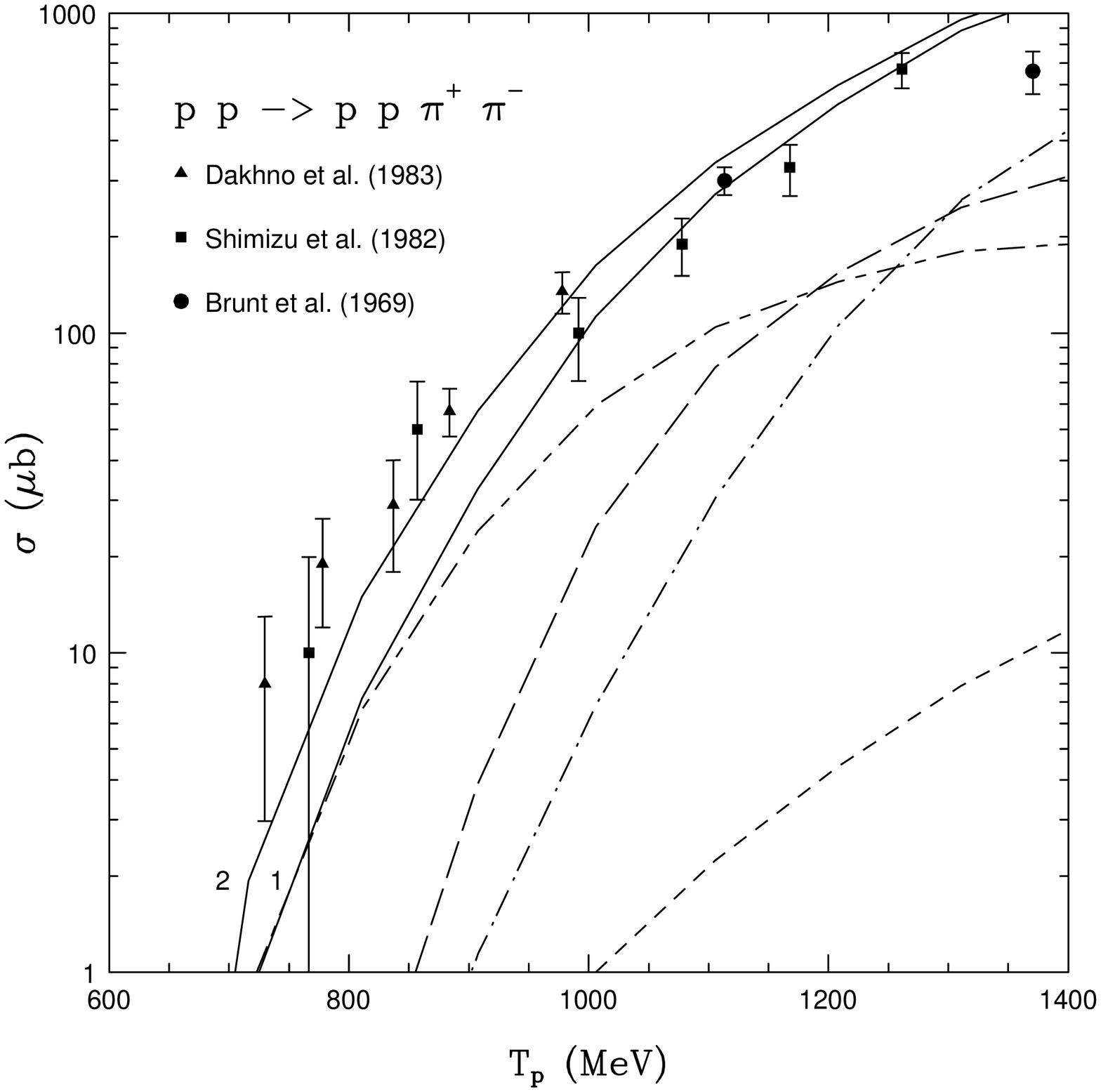,height=5.cm,width=5.cm}}
\end{minipage}
\hfill
\begin{minipage}{.47\linewidth}
\centerline{\psfig{file=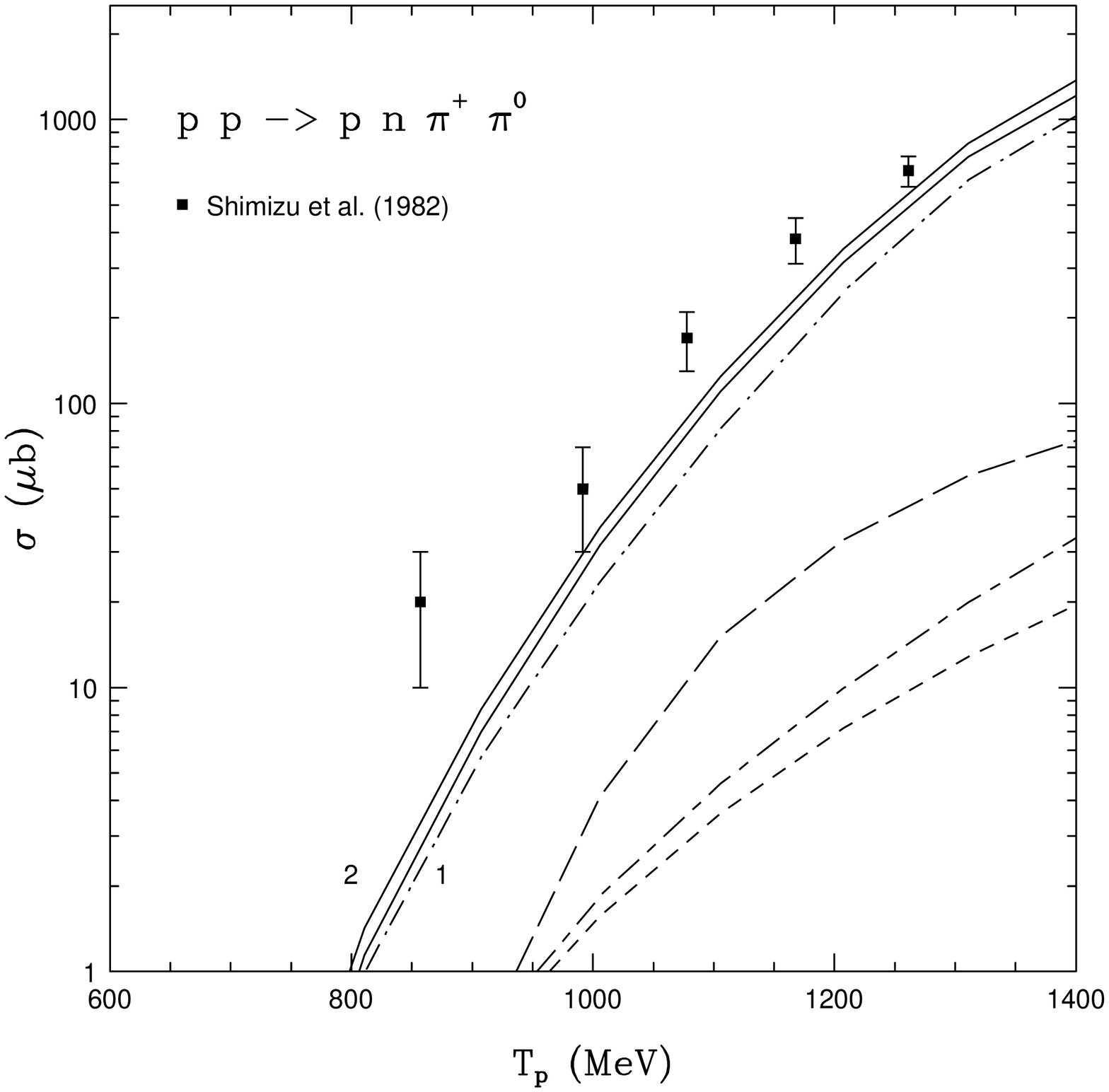,height=5.cm,width=5.cm}}
\end{minipage}
\caption{Total cross sections for two of the channels, as a function of the
incoming proton kinetic energy in lab. frame }
\label{fig6}
\end{figure}

One of the important ingredients in this reaction is that the largest
strength for $N^*$ excitation comes from isoscalar exchange. The 
strength of this transition was obtained from a theoretical analysis
\cite{e2} of the $(\alpha, \alpha')$ reaction on proton targets
exciting the Roper resonance \cite{e3}.

In conclusion we have seen several reactions involving two pions in the
final state. In all of them the $N^*$ resonances play an important
role and we have clarified the links between some resonance
properties and observables in $2 \pi$ production reactions. Further
investigations both theoretical and experimental, extending the work
at higher energies, look also like a fertile land to extend our
knowledge about $N^*$ resonance properties.

\end{document}